\documentclass[12pt,preprint]{aastex}
\usepackage{amssymb, latexsym}
\usepackage{graphicx}
\usepackage{longtable}
\usepackage{amsmath}
\usepackage{natbib}

\begin{document}
\title{Discovery of Three Distant, Cold Brown Dwarfs in the WFC3
  Infrared Spectroscopic Parallels Survey}
\author{D. Masters\altaffilmark{1,2}, P. McCarthy\altaffilmark{2},
 A. J. Burgasser\altaffilmark{3}, N. P. Hathi\altaffilmark{2}, M. Malkan\altaffilmark{4},
	N. R. Ross\altaffilmark{4}, B. Siana\altaffilmark{1},
        C. Scarlata\altaffilmark{5}, A. Henry\altaffilmark{6},
        J. Colbert\altaffilmark{7}, H. Atek\altaffilmark{7},
        M. Rafelski\altaffilmark{8}, H. Teplitz\altaffilmark{8},
        A. Bunker\altaffilmark{9}, A. Dressler\altaffilmark{2}}

\altaffiltext{1}{Department of Physics and Astronomy, University of
  California, Riverside, CA, 92521}
\altaffiltext{2}{Observatories of the Carnegie Institution of
  Washington, Pasadena, CA, 91101}
\altaffiltext{4}{Department of Physics and Astronomy, UCLA, Los
  Angeles, 90095}
\altaffiltext{3}{Center for Astrophysics and Space Science, University
  of California, San Diego, La Jolla, CA 92093}
\altaffiltext{5}{Astronomy Department, University of Minnesota,
  Minneapolis, MN 55455}
\altaffiltext{6}{Department of Physics, University of California,f
  Santa Barbara, CA 93106}
\altaffiltext{7}{Spitzer Science Center, Caltech, Pasadena, CA 91125}
\altaffiltext{8}{Infrared Processing and Analysis Center, Caltech, Pasadena, CA 91125}
\altaffiltext{9}{Department of Physics, University of Oxford, Oxford, UK}

\begin{abstract}

We present the discovery of three late type ($\geq$T4.5) brown dwarfs, including a probable Y dwarf, in the WFC3
Infrared Spectroscopic Parallels (WISP) Survey. We use the G141 grism spectra to determine the spectral types of
the dwarfs and derive distance estimates based on a comparison
with nearby T dwarfs with known parallaxes. 
These are the most distant spectroscopically confirmed T/Y
dwarfs, with the farthest at an estimated distance of $\sim$400
pc.  We compare the number of cold
dwarfs found in the WISP survey with simulations of the brown dwarf mass function.
The number found is generally
consistent with an initial stellar mass function $dN/dM \propto M^{-\alpha}$ with
$\alpha$ = 0.0--0.5, although the identification of a Y dwarf is
somewhat surprising and may be
indicative of either a flatter absolute
magnitude/spectral type relation than previously reported or an upturn
in the number of very late type brown dwarfs in
the observed volume.

\end{abstract}
\maketitle

\section{Introduction}

Understanding the nature and demographics of low-temperature brown dwarfs,
which bridge the gap between dwarf stars and
massive planets, is an important goal of modern astrophysics.
Since the discovery of the first methane dwarf G1229B in 1995
\citep{Nakajima95, Oppenheimer95}, the number of known ultracool
dwarfs of type T and now Y \citep{Cushing11}
has continued to grow, primarily due to deep near-infrared surveys
such as  2MASS (\citealp{Kirkpatrick99, Burgasser02}), SDSS
(\citealp{Strauss99, Chiu06}), UKIDSS
(\citealp{Lodieu07, Chiu08, Burningham10}), CFBDS \citep{Delorme08} 
and WISE (\citealp{Wright10, Kirkpatrick11}). Additional dwarfs have
been identified in proper motion surveys (e.g. \citealp{Kirkpatrick10, Lucas10,
  Scholz11, Liu11}). These surveys have been effective at finding brown dwarfs in the solar neighborhood; however, due to their faintness (a T8 dwarf
has $M_{AB}\sim17$ in H band), probing the larger-scale galactic
distribution of cold dwarfs has not been possible.

Deep surveys in the near-infrared using the Wide Field Camera 3 (WFC3)
on the HST enable the discovery of cold dwarfs at much larger
distances. WFC3 parallel imaging surveys have identified cold dwarfs
based on broad-band colors (e.g. \citealp{Ryan11}), but the lack of
spectroscopy means that the spectral types are uncertain. WFC3 near-infrared slitless grism spectroscopy, in
contrast, permits unambiguous identification and classification of T
dwarfs, whose spectra show
pronounced atmospheric C$\mathrm{H}_{4}$ and $\mathrm{H}_{2}$O 
absorption bands in the near-infrared \citep{Geballe02, Burgasser02}.
Moreover, the WFC3 spectra reach depths that are
unachievable with ground-based surveys. 


Here we present three of the most distant cold brown dwarfs known,
discovered in the WFC3 Infrared Spectroscopic Parallels
(WISP) Survey. These discoveries illustrate that, by enabling
spectroscopic identification of ultracool dwarfs at large distances,
HST-WFC3 grism spectroscopy can begin to probe the galactic
spatial distribution of the ultracool dwarf population and search for substellar
members of the halo population \citep{Burgasser03}.

\section{WISP Survey Overview}

The WISP Survey (\citealp{Atek10}) has obtained
slitless spectra over $0.8$--$1.7\mu$m for more than 700 $\mathrm{arcmin}^{2}$ of sky using
the two infrared grisms installed on the IR channel of Wide Field
Camera 3 (WFC3). While the primary goal of WISP is 
to measure the star formation rate over $0.5<z<2.5$, the wide area and variety of galactic latitudes 
covered by WISP allow for the discovery of late-type dwarfs out to
$\gtrsim1$~kpc. 



WISP is a ``pure parallel'' program, meaning 
that the pointings are determined by 
other observing programs using either the Cosmic Origins Spectrograph (COS) 
or the Space Telescope Imaging Spectrograph (STIS). The WFC3
observations are taken
in a parallel field determined by the offset of the instruments and
the roll angle of the telescope. WISP is divided into a deep, narrow survey for 
parallel targets with more than four orbits of visibility and a shallow, wide 
survey for parallel targets with one to three orbits of visibility. Parallel observing
targets are selected at galactic 
latitudes $|b|>20^{\circ}$\ with a preference for 
longer visibility times. Typical integration times for a 4--5 orbit target are $\sim$5000~sec in G102 and $\sim$2000~sec in G141.
The shallower WISP fields are restricted to the G141 grism, with a typical integration time for a 2-orbit target of $\sim$4000~sec.
Because the wide-shallow survey actually achieves deeper G141 integrations and 
covers a larger area than our deep-narrow survey, faint dwarfs are
more likely to be found in the wide-shallow survey.


\section{Ultracool Dwarfs Discovered in the WISP Survey}

\begin{figure*}
        \centering
	\includegraphics[scale=0.45]{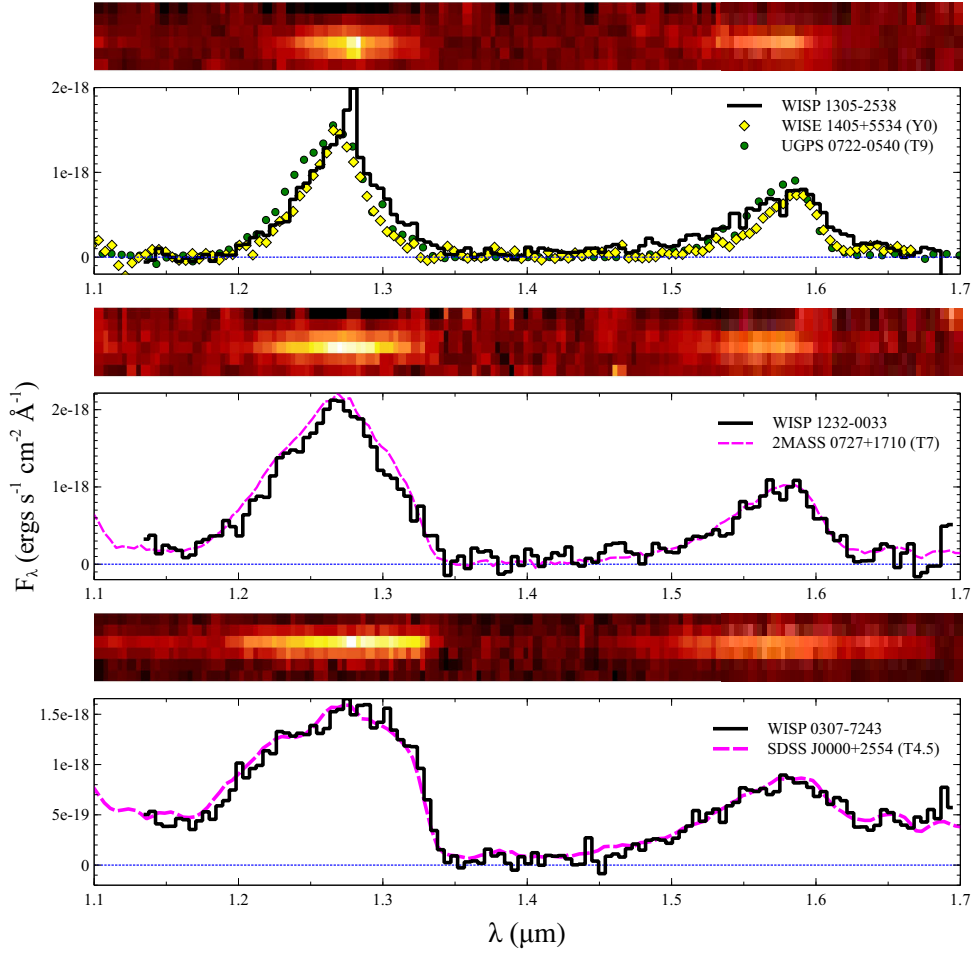}
	\caption{HST-WFC3 G141 1D/2D spectra of the three cold brown dwarfs
          found in the WISP survey, with best-fit template spectra
          overlaid. Prominent pseudo-emission features at roughly
          1.26 and 1.58~$\mu$m are characteristic of T dwarfs and make
          them relatively easy to identify in the 2D grism
          images. While WISP~1232-0033 and WISP~0307-7243 (middle, bottom) are in
          nearly perfect agreement with the best-fit templates,
          WISP~1305-2538 (top) is not perfectly fit by either a T9 or
          Y0 template, although Y0 fits better. The template spectra
          are from the SpeX prism library (\citet{Burgasser10} and
          references therein).}
\label{Fi:dwarfs}
\end{figure*}

Brown dwarfs of
type $\gtrsim$T2 are easily recognizable in H-band
spectroscopy due to the presence of two prominent, broad
pseudo-emission features at roughly 1.26 and 1.58~$\mu$m interspersed
with deep absorption bands of water and methane in their atmospheres.  The
155 G141 grism exposures (each $123''\times136''$) currently obtained by the ongoing WISP survey were systematically scanned
by at least two observers to search for these objects, resulting in the
identification of the three sources shown in
Figure~\ref{Fi:dwarfs}. Their properties are summarized
in Table~1.

\section{Spectral Classification}

We determined the spectral types of the objects by comparing the WFC3 spectra to 69 T
dwarf spectra from the SpeX Prism Spectral Libraries
\citep{Burgasser06} and two additional Y dwarf spectra from
\citet{Cushing11}, which were also obtained with the
HST/WFC3 G141 grism using the same reduction software as the
WISP survey spectra presented here. The WISP spectra and brown dwarf template spectra were
compared using a $\chi^{2}$ statistic. The classifications and associated uncertainties
were then found by computing the weighted mean and deviation from the mean
with a weighting factor of $e^{(\mathrm{min}(\chi^{2})-\chi^{2})^{2}}$. A cutoff
weight of $>$0.1 was used so that the results are not skewed by a large
number of poor fits to early-type T dwarfs (a sampling problem that
arises because the template
sample contains disproportionately more early-type T dwarfs). 

WISP~1232-0033 and WISP~0307-7243 are found to be type T7$\pm$0.6
and T4.5$\pm$0.4, respectively. WISP~1305-2538 is found to be of type T9.5+, and is probably a Y
dwarf. The spectrum of the Y0 dwarf WISE~1405+5534 is shown overlaid in
Figure~\ref{Fi:dwarfs}, in addition to the T9 dwarf
UGPS~0722-0540. While neither template is perfect, the Y0 spectrum is
a better fit.

\section{Distance Determination}

With accurate spectral types derived from the WFC3 spectra, the
distances to the dwarfs can be found by comparing their measured
broad-band magnitudes with those of nearby T dwarfs with known
parallaxes. For this comparison, we converted the WFC3 F140W/F160W
photometry for the WISP dwarfs into MKO
J or H band with color corrections computed using the best-fitting
T dwarf template spectra, applying the same weightings as for the spectral
type determination. Given the inferred MKO J/H magnitudes, the distances were found
using the absolute magnitude/spectral type relations of
\citet{Dupuy12}. 

WISP~1232-0033 (T7) is found to be at 270$\pm$60~pc and WISP~0307-7243 (T4.5) at 400$\pm$60~pc. These are the most
distant T dwarfs currently known, and have vertical scale heights of approximately
240 and 260~pc, respectively. WISP~1305-2538 is more difficult, as
only one Y dwarf has reported a preliminary (and highly uncertain) parallax
(WISE~J154151.66$-$225025.2, $\pi$ = 0$\farcs$35$\pm$0$\farcs$11;
\citealp{Kirkpatrick11}) and the classification scheme for these
objects remains in progress \citep{Cushing11}. We used the known
parallaxes of the T9 UGPS 0722-0540 (0$\farcs$246$\pm$0$\farcs$0033;
Lucas et al. 2010)  and the T9.5 CFBDSIR J145829+101343B
(0$\farcs$340$\pm$0$\farcs$0026; Liu et al. 2011) to infer a distance of 40--60~pc for this source, albeit with considerable uncertainty because of the classification.

\begin{center}
\begin{deluxetable}{lccccccc}
\tabletypesize{\scriptsize}
\setlength{\tabcolsep}{0.03in} 
\tablecolumns{8}
\tablecaption{Cold Brown Dwarfs Discovered in the WISP Survey}
\tablehead{   
  \colhead{Name} &
  \colhead{RA (J2000)} &
  \colhead{DEC (J2000)} &
  \colhead{WFC3 magnitude} &
  \colhead{MKO magnitude} &
  \colhead{Spectral Type} &
  \colhead{Distance} &
  \colhead{Galactic Latitude} \\
  \colhead{} &
  \colhead{(h m s)} &
  \colhead{($^{\circ}$ $'$ $''$)} &
  \colhead{(AB)} &
  \colhead{(AB)} &
  \colhead{} &
  \colhead{(pc)} &
  \colhead{(deg)} 
}
\startdata
WISP 1305-2538  & 13 05 25.51 & -25 38 28.8 &  23.01$\pm$0.04\tablenotemark{b}
& 22.33$\pm$0.04\tablenotemark{d} & T9.5+ &
$\sim$40-60\tablenotemark{a} & +37.1  \\
WISP 1232-0033  & 12 32 42.42 & -00 33 06.7 & 23.34$\pm$0.05\tablenotemark{b} &
22.65$\pm$0.06\tablenotemark{d} & T7$\pm$0.6 &
270$\pm$60 & +63.1  \\
WISP 0307-7243 & 03 07 41.12 & -72 43 57.5 &  22.67$\pm$0.02\tablenotemark{c} &
22.36$\pm$0.02\tablenotemark{e} & T4.5$\pm0.4$ & $400\pm60$ &
-40.9 
\enddata
\tablenotetext{a}{The interval arises from
  classification uncertainty and the current lack of quality
  parallaxes for Y dwarfs.}
\tablenotetext{b}{F140W}
\tablenotetext{c}{F160W}
\tablenotetext{d}{J band}
\tablenotetext{e}{H band}
\end{deluxetable}
\end{center}
\normalsize

\section{Space Density}

The dwarfs discovered in the WISP survey provide the first
opportunity to examine the mass function and spatial distribution of these late-type dwarfs in
a deep field (see \citealp{Pirzkal09} for an equivalent study of late
M and L dwarfs). To assess what constraints our small sample makes on
these statistics, we computed the expected suface densities of T
dwarfs as a function of spectral type in the WISP survey fields as follows.
\begin{enumerate}
\item{Volume densities for the cold dwarf subtypes are computed
    using the methods outlined in \citet{Burgasser04, Burgasser07}. We assume a power-law stellar mass function
    $dN/dM~\sim~M^{-\alpha}$ with $\alpha$ varying
    from -1 to 1, normalized to agree with
    measured field values in the 0.09-0.10 $\mathrm{M}_{\odot}$ range
    (0.0055 pc$^{-3}$; \citealp{Reid99, Chabrier01}).
    We then assumed an age distribution consistent with a constant star
    formation rate over 10 Gyr, and computed the luminosity
    distribution for the dwarfs using the evolutionary models of
    \citet{Burrows97}. Finally, we use empirical bolometric
    correction/spectral type \citep{Liu10} and absolute
    magnitude/spectral type \citep{Dupuy12} relations to convert the
    luminosity function to a spectral type surface density distribution.}
\item{The expected surface density of dwarfs per WISP
    $123''\times136''$ pointing is then
    found by computing the limiting magnitude, and thus
    the maximum distance, for which each subtype would be
    spectroscopically detected\footnote[1]{We require S/N=10 
in this analysis, an approximation based on simple tests with the
WFC3 spectra. However, changing the S/N requirement can
significantly influence the expected surface density of dwarfs.}
    in a given field. The online WFC3 exposure time calculator is
    used for this analysis. Corrections to the derived limiting magnitudes are applied to
    account for Malmquist bias. We assumed as a first order
    approximation that 10\% of each pointing is lost due to
    overlap, edge effects, etc. as estimated from the galaxy extractions.}
\item{The effective volumes for each field/subtype are
    found and multiplied by the space densities per subtype. The
    results for all
    the WISP fields are
    added, yielding a surface density for each subtype for the entire
    survey (Figure~2).}

\end{enumerate}

The expected surface density of each spectral type across the entire
WISP survey is summarized in Figure~2 for different assumed
power-law slopes in the low-mass stellar mass function. The number of dwarfs found is
roughly consistent with expectations; in particular, the
fact that we find no early type ($<$T4) dwarfs, despite their
higher intrinsic luminosities, is consistent
with their rarity \citep{Burgasser07}. The fact that we identified
three T dwarfs is consistent with other field mass function estimates
based on wide, shallow surveys of T dwarfs (e.g. \citealp{Metchev08,
  Burningham10}). However, finding
a T9.5/Y0 dwarf is somewhat surprising due to their faintness. Possible
explanations for this include (1) a 
flatter absolute magnitude/spectral type relation than preliminary
estimates indicate \citep{Cushing11, Kirkpatrick11}, (2) a sudden
upturn in the space density of very late-type dwarfs, possibly
associated with a nearby free-floating planetary population
\citep{Sumi11} or a larger proportion of old halo brown dwarfs at
higher scale heights (for this simulation, a 1:400 halo fraction was
assumed based on \citealp{Digby03}), (3) a fortuitous discovery, or
(4) some combination of the above. We note that WISP~1305-2538 is
4$\farcm$2 from the very low-mass triple Kelu~1ABC at 18.7$\pm$0.7~pc
(\citealp{Dahn02, 2005ApJ...634..616L, 2008arXiv0811.0556S}).  With a 5000~AU
projected separation, this matching is unlikely \citep{Dhital10}, but
second-epoch imaging would allow us to affirm or refute the existence
of a widely-separated, very low-mass quadruple. Additionally,
follow-up proper motion measurements of all three new dwarfs would
help establish whether they are galactic halo members.

\section{Summary}

We have presented three distant T/Y dwarfs spectroscopically confirmed
by the WISP
Survey. These discoveries show the power of HST-WFC3 slitless grism
spectroscopy to simultaneously find these objects at much larger distances than
has been possible with ground-based surveys and yield
accurate spectral types, making surveys such as WISP powerful new
probes of the population of cold brown
dwarfs in the galaxy.

\begin{figure*}[t]
        \centering
	\includegraphics[scale=0.55]{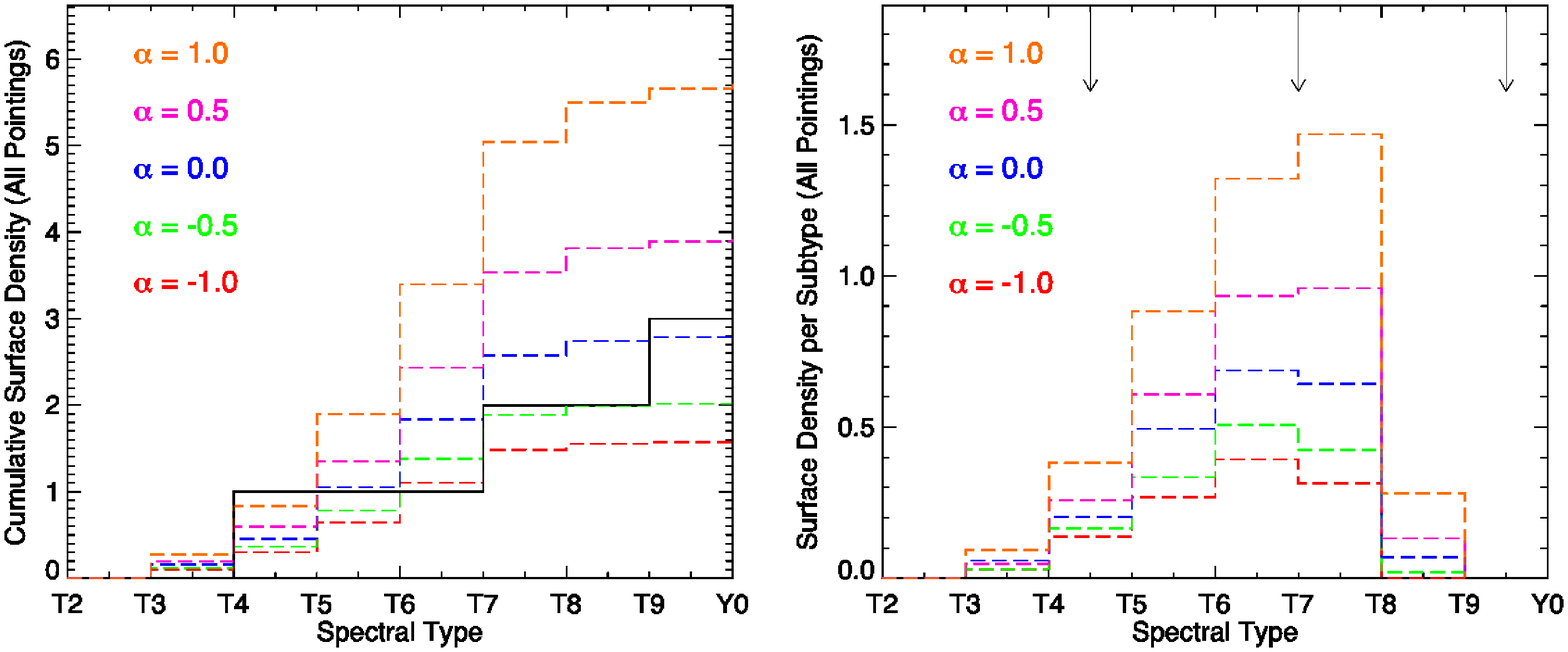}
	\caption{{\bf Left:} Expected cumulative surface density of
          late-type dwarfs across all WISP fields for different values
          of $\alpha$ in the power-law stellar mass function
          $dN/dM~\sim~M^{-\alpha}$, from the simulations described in
          \S6. The total number found (three, black line) is
          generally consistent with $\alpha$=0.0--0.5. {\bf Right:}
          The expected surface density of each spectral type across
          the entire WISP survey. The spectral types of the three
          dwarfs found are indicated with downward arrows. The discovery of a
          T9.5/Y0 dwarf is somewhat surprising; see \S6 for a
          discussion of possible explanations.}
\label{Fi:surfacedensity}
\end{figure*}

\acknowledgments
We thank an anonymous referee for helpful comments that improved this letter. This research has benefitted from the SpeX Prism Spectral Libraries,
maintained by Adam Burgasser at
http://www.browndwarfs.org/spexprism. We thank Michael Cushing for
providing Y0 spectra from the WISE survey. 

This work has been
supported by the Carnegie Observatories Graduate Research Fellowship.

\pagebreak
\bibliographystyle{plainnat}

\end{document}